\documentclass{article}

\usepackage{arxiv}

\usepackage[utf8]{inputenc} 
\usepackage[T1]{fontenc}    
\usepackage{hyperref}       
\usepackage{url}            
\usepackage{booktabs}       
\usepackage{amsfonts}       
\usepackage{nicefrac}       
\usepackage{microtype}      
\usepackage{lipsum}		
\usepackage{graphicx}
\usepackage{natbib}
\usepackage{doi}

\usepackage{amsmath,amssymb,amsfonts}
\usepackage{textcomp}
\usepackage{algorithmic}
\usepackage{xcolor}
\usepackage{comment}
\usepackage{cleveref}
\usepackage{colortbl}
\usepackage[nolist]{acronym}
\usepackage{float}
\usepackage{booktabs}

\title{Construction and Control of Validated Highly Configurable Multi-Physics Building Models for Multi-Energy System Analysis in a Co-Simulation Setup}

\date{}

\author{ \href{https://orcid.org/0009-0005-0067-8019}{\includegraphics[scale=0.06]{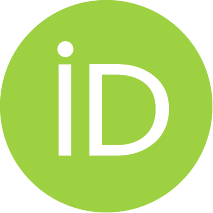}\hspace{1mm}Haozhen~Cheng}\thanks{This work was conducted within the framework of the Helmholtz Program Energy System Design (ESD) and is partially funded under the project “Helmholtz platform for the design of robust energy systems and their supply chains” (RESUR).} \\
	Institute for Automation and Applied Informatics\\
	Karlsruhe Institute of Technology\\
	76131 Karlsruhe, Germany \\
	\texttt{haozhen.cheng@kit.edu} \\
	\And
	\href{https://orcid.org/0000-0001-5284-166X}{\includegraphics[scale=0.06]{orcid.pdf}\hspace{1mm}Jan~Stock} \\
	Forschungszentrum Jülich\\
	52428 Jülich, Germany \\
	\texttt{j.stock@fz-juelich.de} \\
        \And
	\href{https://orcid.org/0000-0001-5399-2363}{\includegraphics[scale=0.06]{orcid.pdf}\hspace{1mm}André~Xhonneux} \\
	Forschungszentrum Jülich\\
	52428 Jülich, Germany \\
	\texttt{a.xhonneux@fz-juelich.de} \\
        \And
	\href{https://orcid.org/0000-0002-1463-7606}{\includegraphics[scale=0.06]{orcid.pdf}\hspace{1mm}Hüseyin K.~Çakmak} \\
	Institute for Automation and Applied Informatics\\
	Karlsruhe Institute of Technology\\
	76131 Karlsruhe, Germany \\
	\texttt{hueseyin.cakmak@kit.edu} \\
        \And
	\href{https://orcid.org/0000-0002-3572-9083}{\includegraphics[scale=0.06]{orcid.pdf}\hspace{1mm}Veit~Hagenmeyer} \\
	Institute for Automation and Applied Informatics\\
	Karlsruhe Institute of Technology\\
	76131 Karlsruhe, Germany \\
	\texttt{veit.hagenmeyer@kit.edu} \\
}

\renewcommand{\shorttitle}{Validated Multi-Physics Building Models for Multi-Energy System Analysis}

\hypersetup{
pdftitle={Construction and Control of Validated Highly Configurable Multi-Physics Building Models for Multi-Energy System Analysis in a Co-Simulation Setup},
pdfsubject={eess.SY},
pdfauthor={Haozhen~Cheng, Jan~Stock, André~Xhonneux, Hüseyin K.~Çakmak, Veit~Hagenmeyer},
pdfkeywords={Building, Modelica, Heat grid, Validation, Co-Simulation},
}

\begin{document}

\begin{acronym}
    \acro{KIT}{Karlsruhe Institute of Technology}
    \acro{FZJ}{Forschungszentrum Jülich}
    \acro{DH}{District Heating}
    \acro{TABULA}{Typology Approach for Building Stock Energy Assessment}
    \acro{DHW}{Domestic Hot Water}
    \acro{HIU}{Heat Interface Unit}
    \acro{HVAC}{Heating, Ventilation and Air Conditioning}
    \acro{FMI}{Functional Mock-up Interface}
    \acro{FMU}{Functional Mock-up Unit}
    \acro{LPG}{LoadProfileGenerator} 
\end{acronym}

\maketitle

\begin{abstract}
    Improving energy efficiency by monitoring system behavior and predicting future energy scenarios in light of increased penetration of renewable energy sources are becoming increasingly important, especially for energy systems that distribute and provide heat. On this background, digital twins of cities become paramount in advancing urban energy system planning and infrastructure management. The use of recorded energy data from sensors in district digital twins in collaborative co-simulation platforms is a promising way to analyze detailed system behavior and estimate future scenarios. However, the development and coupling of multi-physics energy system models need to be validated before they can be used for further in-depth analyses. In the present paper, a new multi-physics/-modal and highly configurable building model is presented. Its accuracy and reliability are validated by comparison with data from the TABULA project, ensuring its relevance and applicability to real-world scenarios. The modularity and flexibility with regard to the system configurability of the developed building model is evaluated on various real building types. In addition, the applicability of the building model in a multi-energy system is highlighted by implementing the model in a collaborative co-simulation setup and by coupling it to a district heating grid model in yearly co-simulations. The simulation results for the proposed multi-physical/-modal building modeling concept show a very high level of agreement compared to published reference building data and can therefore be used individually as flexible and modular building models including both thermal and electrical systems for future sector-coupled energy system analyses in view of sustainability.
\end{abstract}


\section{Introduction}
\label{sec:Introduction}
Digital twins in the energy system domain offer multiple benefits, as they can improve operational efficiency for sustainability and help decision-making by enabling the evaluation of various scenarios based on a virtual representation of physical assets fed by real-time monitoring data. Further benefits are the possibility for planning extensions of existing energy systems, e.g., for incorporating renewable energy sources or storage systems. Additionally, they can help to detect existing issues in energy systems and to develop optimization strategies. For the design of energy digital twins of districts, it is essential to develop models of building and energy grids considering aspects such as model validation, definition of model interfaces, and enabling the coupling of these models using advanced computing architectures for the evaluation using co-simulation techniques. 

As a motivation for developing a district energy twin, an ongoing project at Energy Lab \citep{hagenmeyer2016} at \ac{KIT} with investors, architects, urban planners, and energy network operators has been established. The target area comprises nearly 30 buildings of various types (single and multifamily homes), a 20kV transformer for the electricity supply of the low voltage grid, and a CHP (combined heat and power) with a heat grid. All buildings will be equipped with smart meters for electricity and heat demand, thus enabling the transfer of all measurement data to the Energy Lab for storage and further data analysis. The data will be available for developing new control algorithms, and forecasting \citep{kovacevic2024}, and also serve as input for the district digital twin.

On this background, the authors are investigating new approaches to model and simulate the coupling for multi-physics-based energy system models in view of sustainability. The focus is on the development of flexible and modular building models and heat grids, together with their interconnection interfaces for interconnected co-simulation.

\subsection{Related work}

\ac{DH} systems have been extensively studied in the past using different modeling approaches and following various objectives. Brown et al. gave an overview of approaches developed and highlight several commercial tools \citep{brown2022}.
Simulation approaches are available in different programming languages and are mainly developed for dedicated applications in the field of \ac{DH} simulation. The design of \ac{DH} models was addressed in \citep{fuchs2016} and \citep{schweiger2017} using Modelica (modelica.org) models, while Stock et al. \citep{stock2023} developed a tool based on Python. The integration of a waste heat source into an existing network structure was demonstrated in \citep{lohmeier2020} using Modelica models with multiple heat sources. Using dynamic simulation models, Capone et al. \citep{capone2023} investigated the optimal placement of central heat pumps in a high-temperature \ac{DH} system. While \ac{DH} system simulations are widely represented, the detailed simulation of existing connections between network structure and buildings, e.g., by coupling existing \ac{DH} grids and supplied buildings, each of which is modelled in an individual simulation environment, is rarely investigated by highly detailed models.

As for previous works on building modeling, Johari et al. \citep{johari2022} evaluated several buildings in different configurations using different building energy models \citep{ciccozzi2023} and modeling tools such as IDA ICE (www.equa.se/de/ida-ice), TRNSYS (www.trnsys.com), and EnergyPlus (energyplus.net), and analyzed the applicability of these tools for energy analysis of different types of buildings. They focused on comparisons between tools, yet lacked experience with the Modelica language. Based on three types of buildings presented from the \ac{TABULA} \citep{tabula2016} project i.e., single house, terrace house and multifamily house, Bruno et al. \citep{bruno2016} analyzed the simplified thermodynamic behavior of the buildings using the TRNSYS tool. They focused on analyzing simplified building envelope models and individual building heat loads without the connection between buildings and energy grids. Although Modelica is not a mainstream tool in modeling building energy demand \citep{alfalouji2023}, Perera et al. \citep{perera2016} developed a multilayer building heating model in MATLAB (mathworks.com) and Modelica environments and showed that Modelica models are more robust.

Due to the lack of freedom to control and adjust the building's internal components, as well as controllers, to the best knowledge of the authors, no study has observed and analyzed the building's internal variables over time.  
On this background, the present paper introduces a new building modeling concept, whose performance is demonstrated with a coupled heat grid in a co-simulation environment.

\subsection{Contributions and outline}

The contribution of this paper is a new validated multi-physics/-modal and flexible configurable building model with interfaces for coupling with heat grids for building sustainability analysis, aiming to the development of energy digital twins of districts. The building model is a generic but parameterizable residential building model using the Modelica language and equipped with interfaces to interact with other energy grids. With different parameter configurations and input data, in combination with Python code, the model can be used to generate any number of different residential buildings that are individually configurable regarding their internal energy (supply) systems. 
Therefore, the proposed building model can easily be used for multi-domain co-simulation in a specific simulation platform with energy systems such as power, heat, or gas grids as well as sectoral coupling elements, in order to lay the foundation for future analysis of the regional energy system in special or extreme scenarios such as natural gas shortages or extreme weather conditions like heat waves or any other disruptive events.

The remainder of the present paper is structured as follows:
In \Cref{sec:Modular building model}, a modular and flexible building model comprising envelope, equipment, interfaces, and control is presented. \Cref{sec:District Heating System} presents a model for a heating system that is coupled with the building models in \Cref{sec:Model and Simulation Coupling} using a distributed co-simulation framework. In \Cref{sec:Evaluation} the authors present the results of the model validation for various building types as well as the coupled simulation in a co-simulation setup with a discussion of the results. The paper is concluded with a brief outlook on the future work in \Cref{sec:Conclusion and Outlook}.

\section{Modular building model}
\label{sec:Modular building model}
\begin{figure}[htb]
    \centerline{\includegraphics[width=0.65\columnwidth]{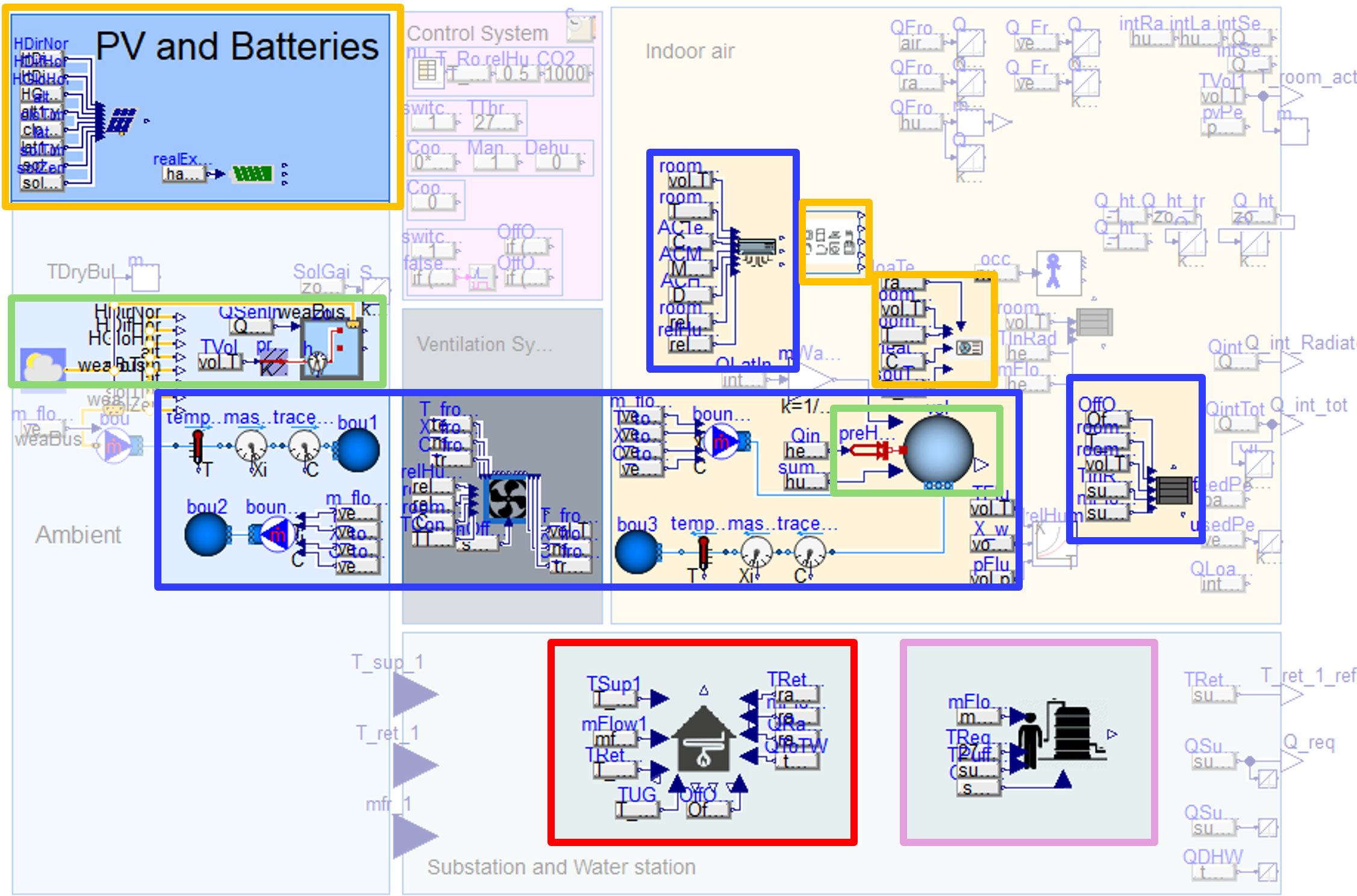}}
    \caption{Modular building model built in \textit{Dymola} using the \textit{Modelica} language. The building envelope model structure is labeled with green boxes, the \ac{HVAC} system with blue boxes, the \ac{HIU} with a red box, the electrical equipments with orange boxes, and the \ac{DHW} station with a pink box.}
    \label{fig: House model}
\end{figure}

In the field of building energy, reliable modeling of buildings is crucial to analyze their energy consumption and for planning their adaptation to energy systems. A physical modeling language that meets these requirements is Modelica, an object-oriented physical modeling language. Although it is an equation-based modeling method like many modeling tools (Simulink etc.), due to the acausal nature of its model, the models created by Modelica are more concise and clear at a glance in more complex systems, especially when it is necessary to understand the behavior and interaction of the model in a complex and multi-domain system. In the present study, the Modelica modeling language is utilized for performing modeling work in the Dymola modeling and simulation environment \citep{Dymola3ds2024}. \Cref{fig: House model} depicts the schematic diagram of the developed novel building model with several modules in Dymola, showing its modular nature. The following sections will introduce each component of the proposed building model.

\subsection{Building envelope model}
\label{subsec:Building envelope model}

\begin{figure}[htb]
    \centerline{\includegraphics[width=0.3\columnwidth]{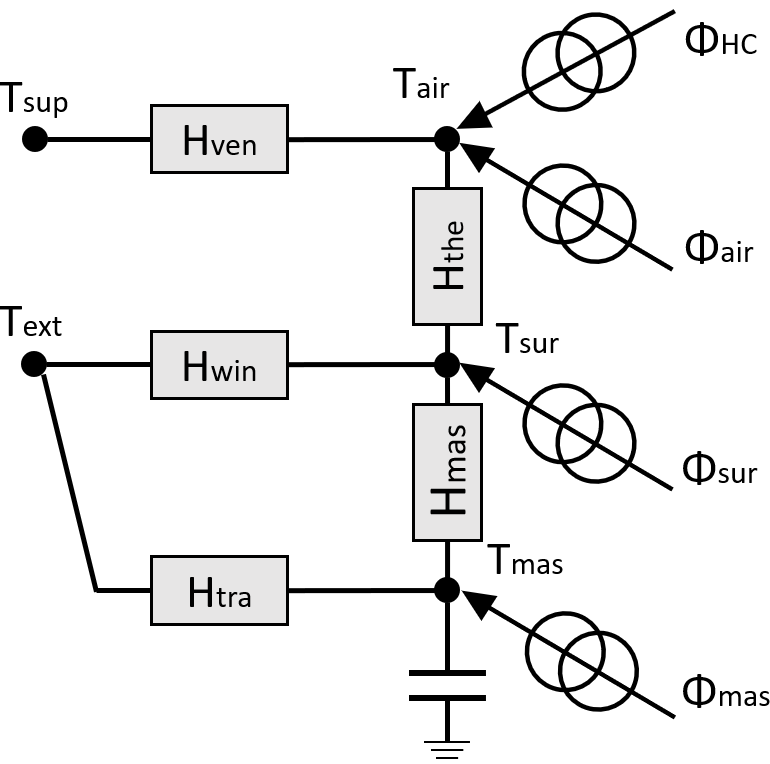}}
    \caption{Electrical schematic diagram of the 5R1C building envelope model \citep{bruno2016}.}
    \label{fig: 5R1C electrical scheme model}
\end{figure}

In order to calculate a building's heating or cooling demands, it is important to use a reasonable building envelope model. In the Modelica \textit{Buildings} library \citep{WetterZuoNouiduiPang2014}, the developers of the library have established a simple, clear and easy-to-understand mathematical model (see \Cref{fig: 5R1C electrical scheme model}), namely the \textit{5R1C} building envelope model, which is described in the EN ISO 13790:2008 standard \citep{eniso2008}. As shown in \Cref{fig: 5R1C electrical scheme model}, this model is analogous to the resistance and capacitance in the circuit model and simultaneously considers characteristics of heat conduction and heat capacity in buildings through 5 \textit{conductance} elements representing the heat conduction characteristics, 5 \textit{voltage nodes} representing the temperature, and 1 \textit{capacitance} element representing the building heat capacity of the entire zone.

The indoor part of the building model consists of three temperature nodes, namely the indoor air temperature $T_{air}$, the internal surface temperature of the envelope $T_{sur}$ and the area mass temperature $T_{mas}$, which are thermally connected to two temperature nodes located outdoors -- with $T_{sup}$ representing the supply air temperature and $T_{ext}$ representing the outdoor air temperature -- through various thermal conductance elements. Among them, the window heat transfer $H_{win}$ and the heat transfer of the opaque component $H_{op}$ (divided into $H_{tra}$ and $H_{mas}$) determine the indoor thermal transmission characteristics with outside. $H_{mas}$ is the thermal conductance between the surface and the mass nodes and is defined as:
\begin{equation}
\label{eq: Hmas}
H_{mas} = h_{ms} \cdot f_{ms} \cdot A_{f}
\end{equation}
where $h_{ms}$ is the heat transfer coefficient between the two nodes, with a fixed value of  9.1$W/(m^{2}K)$, $f_{ms}$ is a correction factor, which can be assumed as 2.5 for light and medium building constructions, and 3 for heavy, and $A_{f}$ is the floor area. Since the transmission coefficients for opaque and glazing elements $H_{op}$ and $H_{win}$ are defined by their respective U-values and areas, the expression of $H_{tra}$ can be derived:
\begin{equation}
H_{tra} = \frac{1}{\frac{1}{H_{op}} - \frac{1}{H_{mas}}}
\end{equation}

In addition, the heat transfer characteristics between indoor air and inner surface are expressed by $H_{the} = h_{as} \cdot A_{tot}$, where $h_{as}$ is the heat transfer coefficient between the air node and the surface node, also with a fixed value of 3.45$W/(m^{2}K)$, and $A_{tot}$ is the whole area of all surfaces facing the air-conditioned building area which is often chosen as a fixed multiple ($rat_{sur}$) of the floor area $A_{f}$, such as 4.5. Since different buildings are studied in the present paper, the authors make the following estimate of this area to obtain a calculation result as accurate as possible:
\begin{equation}
A_{tot} = A_{roof}+A_{wall}+(2n-1)A_{floor}+A_{window}
\end{equation}
where $n$ is the number of layers of the building.

In order to make the modeling results as correct as possible, the structure of the model and many parameters need to be reasonably designed. The European project \ac{TABULA} \citep{tabula2016} aims to enhance energy efficiency in the residential building sector by developing a comprehensive framework for evaluating and improving the energy performance of building stocks across Europe. In the present paper, the authors select two different single-family houses (DE.N.SFH.05.Gen and DE.N.SFH.07.Gen) from \ac{TABULA} as reference objects, and each building is divided into two different refurbishment states, thus four different parameter configurations in total. The parameter configurations of the four selected buildings in \ac{TABULA} are shown in \Cref{tab: Buildings parameters}. A building number ending in ".001" represents a building in the existing state, while the ending ".002" represents a building state after usual refurbishment. Based on this, the authors configure the same parameters for the Modelica models in the present paper and added, deleted and adjusted some parameters in Dymola according to the differences between the \ac{TABULA} and 5R1C models as detailed in the following subsections.

\begin{table*}[htb]
\centering
    \caption{The parameters of the four building models in \textit{\ac{TABULA}} and their corresponding parameter configuration in \textit{Dymola}}
    \setlength{\fboxrule}{0pt}
    \fbox{\includegraphics[width=1\linewidth,trim={0.0cm 0.0cm 0.0cm 0.0cm},clip]{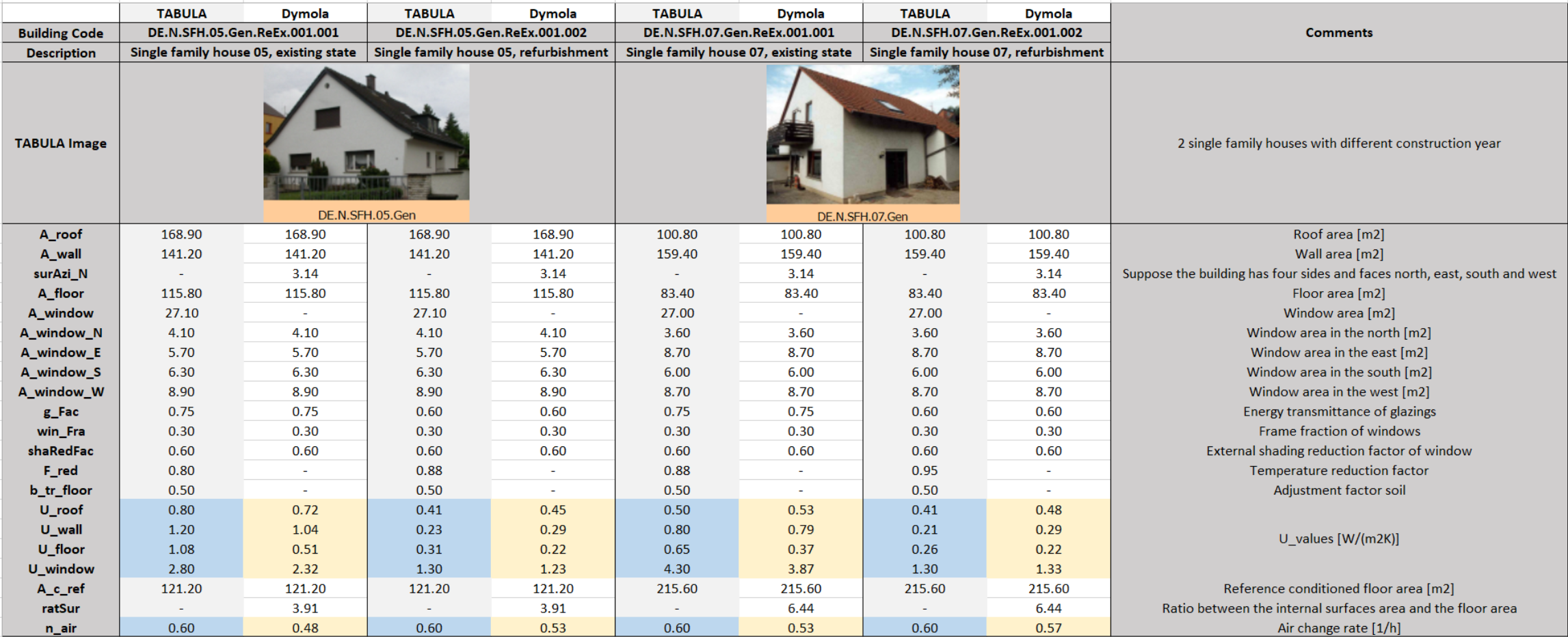}}
    \label{tab: Buildings parameters}
\end{table*}

\subsubsection{Temperature reduction factor}
In the \ac{TABULA} project, non-uniform heating is taken into account, e.g., internal temperatures are reduced during intermittent heating periods. This temperature reduction effect is influenced by building insulation and is more pronounced in poorly insulated buildings. Therefore, \ac{TABULA} added the temperature reduction factor $F_{red}$ to the calculation of the heat transfer performance of the building, as shown in \Cref{tab: Buildings parameters}.

\subsubsection{Thermal bridging}
Since, according to the \ac{TABULA} method, the thermal envelope area of a building is calculated based on the external dimensions, heat losses are generally slightly overestimated at joints (e.g., masonry edges) of building elements made of homogeneous materials. During the calculation of the building transmission heat transfer, the corresponding additional losses are added and their U-values $\Delta U_{tbr}$ are assumed to be 0.1$W/(m^{2}K)$. Thus, the total heat transfer coefficient $H_{tr}$ is calculated by the following equation:
\begin{equation}
H_{tr} = \sum_{i} b_{tr,i}A_{env,i}U_{i} + (\sum_{i} A_{env,i})\Delta U_{tbr}
\end{equation}
where $b_{tr,i}$ is the adjustment factor, which is 1 for roof, wall, and window and 0.5 for floor. $A_{env,i}$ is the sum of all surface areas and $U_{i}$ is the U-value of the surfaces. The equivalent U-value for each surface is calculated in the Modelica model by the following equation:
\begin{equation}
U_{i,eq} = b_{tr,i}U_{i} + \Delta U_{tbr}
\end{equation}

Since \ac{TABULA} also considers the temperature reduction factor, the U-value should continue to be rewritten as:
\begin{equation}
U_{i,eq,red} = (b_{tr,i}U_{i} + \Delta U_{tbr}) F_{red}
\end{equation}
and thus $H_{tr}$ becomes
\begin{equation}
H_{tr} = \sum_{i} A_{env,i}U_{i,eq,red}
\end{equation}

After such a transformation, the result of the parameter configuration in Modelica is shown in the yellow background cell in \Cref{tab: Buildings parameters} (see rows $U_{roof}$ to $U_{window}$).

\subsubsection{Air change rate}
Ventilation systems are mainly used to control indoor air quality, temperature, and relative humidity. In the \ac{TABULA} project, the air exchange rate is assumed to be 0.6$\frac{1}{h}$, which is defined in
\begin{equation}
n_{air,rate} = n_{air,use} + n_{air,infiltration}
\end{equation}
as a constant value for the whole year. So the air change rate configured for the model in the present paper is also matched to it and also takes into account the temperature reduction factor:
\begin{equation}
n_{air,rate,red} = n_{air,rate} \cdot F_{red}
\end{equation}
which is shown in \Cref{tab: Buildings parameters} (named $n\_air$).

\subsection{Building equipment}
\label{subsec:House equipment}
In order to ensure the simplicity of the model and the speed of simulation, while also trying to ensure the appropriateness of the model and focusing on the study of building thermal requirements, the authors add new components to the model.

\subsubsection{\ac{HVAC} system}
\label{subsec:HVAC System}
The \ac{HVAC} system represents critical components in maintaining indoor environmental quality and comfort in residential, commercial, and industrial buildings. Among them, the ventilation fan controls the relative humidity and $CO_{2}$ concentration in the building during the simulation, and performs temperature control when necessary, such as when the room overheats in the summer; radiators transfer heat to indoor spaces through convection and radiation, and can be powered by various heat sources; air conditioners can adjust the temperature and humidity of indoor temperature by consuming electricity. The corresponding positions of the three components are shown in \Cref{fig: House model}.

In the model designed in the present paper, the radiator can control the mass flow rate of hot water from the buffer tank into the radiator by detecting the difference between the temperature of the room and the set temperature.

\subsubsection{Electrical appliances}
\label{subsubsec:Electrical appliances}
\begin{figure}[htb]
    \centerline{\includegraphics[width=0.8\columnwidth]{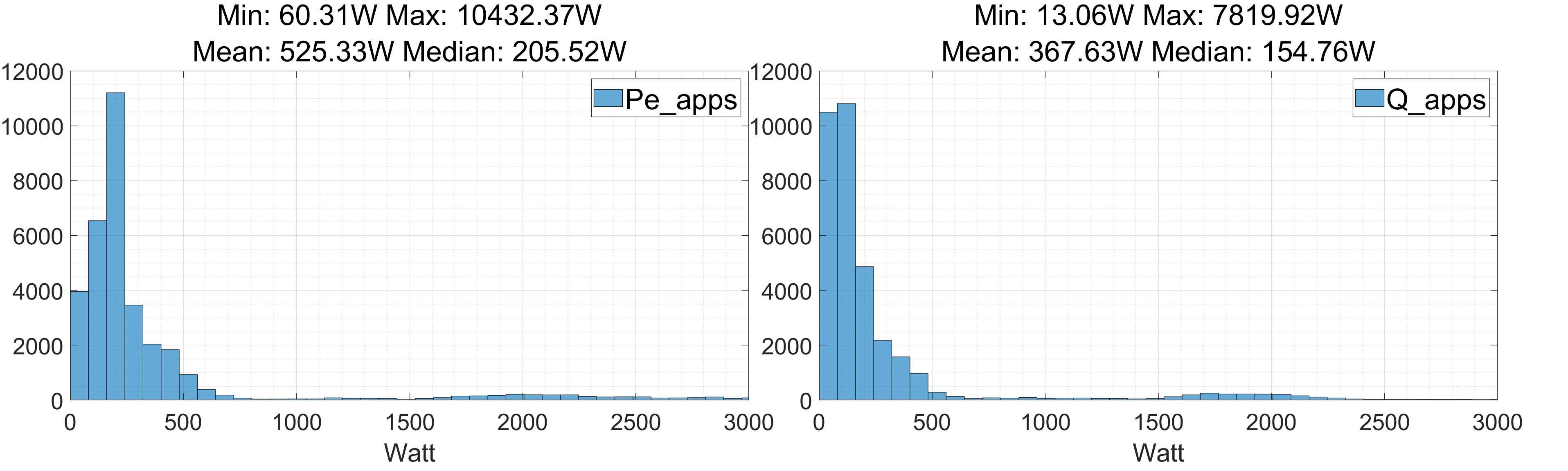}}
    \caption{Distribution of power consumption (Pe) and heat production (Q) of electrical appliances, total per year 4602kWh electricity and 3220kWh heat.}
    \label{fig: APP}
\end{figure}

In the present paper, the authors assume that there are three occupants in each building, and use the \textit{\ac{LPG}} software \citep{lpg2022} to calculate four identical power consumption profiles (see the profiles after histogram transformation in \Cref{fig: APP}) assigned to the four buildings. Further, the authors assume -- based on the information of numerous items in the well-known furniture and appliance websites and available open data -- the ratio of heat generation and electrical energy consumption of each appliance given in \Cref{tab:apps} that is used for calculating the heat from the appliances in each building. In this way, the heat gain inside the four buildings from the appliances is identical.

\begin{table}[htb]
    \centering
    \caption{Types of electrical appliances calculated and their ratios of thermal and electrical power}
    \label{tab:apps}
    \resizebox{0.4\columnwidth}{!}{ 
    \begin{tabular}{lc}
    \hline
    \rowcolor[HTML]{EFEFEF} 
    \textbf{Type} & \textbf{Heat-to-Electricity Ratio}  \\ 
    \hline
    {Light} & 0.95 \\
    {Stove} & 0.99 \\
    {Coffee Machine} & 0.1 \\
    {Toaster} & 0.98 \\
    {Electric Kettle} & 0,9 \\
    {Entertainment} & 0.5 \\
    {Hood} & 0.01 \\
    {Oven} & 0.98 \\
    {Dryer} & 0.95 \\
    {Fridge} & 1 \\
    {Dishwasher} & 0.05 \\
    {Microwave} & 0.3 \\
    {Washing machine} & 0.01 \\
    {Hair dryer} & 0.9 \\
    {Vehicle} & 0 \\
    \bottomrule
    \end{tabular}%
    }
\end{table}

\subsubsection{Domestic hot water supply}
\label{subsubsec:Domestic hot water supply}
\ac{DHW} supply is an important utility, often used for bathing, cleaning, and cooking. The authors assume that each building has an independent thermostatic storage tank for \ac{DHW}. In case of a demand for hot water, a heat exchanger transfers heat from the \ac{DH} water to the domestic water supply. The model established in the present paper allows users to choose whether to turn on or off \ac{DHW} station based on computing needs. Since this paper focuses on the validation of the model in the context of building indoor temperature control, the \ac{DHW} module is deactivated and is not considered further.

\subsubsection{Further modules}
\label{subsubsec:Other modules}
Many other modules are developed in the present building model but will not be discussed in detail in the present paper, such as the photovoltaic module, the battery module, and the heat pump module. Their presence in this building model provides a high degree of flexibility and versatility while placing great challenges on the model's control.

\subsection{Building and district heating grid interface}
\label{subsec:Building and district heating grid interface}
\begin{figure}[htb]
    \centerline{\includegraphics[width=0.6\columnwidth]{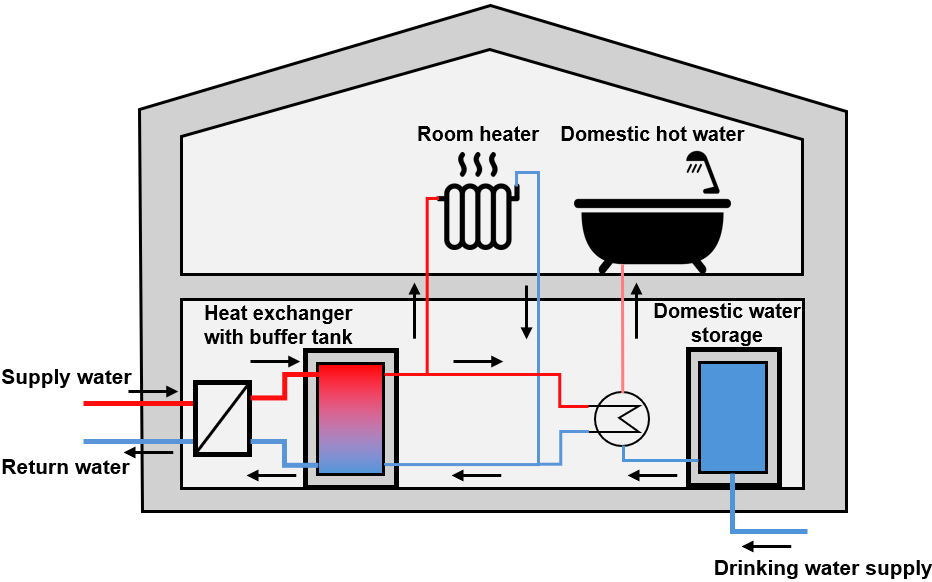}}
    \caption{\ac{HIU} for room heating and \ac{DHW}.}
    \label{fig: HIU}
\end{figure}

The interface between the building and the \ac{DH} grid is called a \ac{DH} substation or \ac{HIU}. HIUs can generally be divided into two categories: those used for room heating only and those used for room heating and the production of \ac{DHW} \citep{cibse2011}, as shown in \Cref{fig: HIU}. In addition, in terms of heat supply, \ac{HIU} can also be divided into direct and indirect \ac{HIU}. In a direct \ac{HIU}, hot water from the \ac{DH} grid enters the building's heating and hot water systems directly through pipes. Its system design is relatively simple, but there may be risks of water supply contamination and pressure fluctuations. On the contrary, indirect \ac{HIU} transfers heat from the \ac{DH} grid to the building's heating or hot water systems via heat exchangers, which provide isolation of the water system, and water quality is ensured. In addition, a buffer tank is modeled in \ac{HIU}. The main function of the buffer tank is to store thermal energy. It allows the system to buffer heat supply and smooth demand changes, this ensures that there is always a reserve of hot water or thermal energy available, even during peak usage periods. Details will be discussed in \Cref{subsec:Interface on the Building Side}.

\subsection{Control of building model}
\label{subsec:Control of building model}
This section describes the control of some components and variables during simulation.

\subsubsection{Set point of room temperature}
\label{subsubsec:Set point of room temperature}
Tenants in Germany benefit from laws regulating the heating season, which runs from October 1 to April 30 \citep{heating2024}. Nevertheless, since the heating season is set to 222 days in \ac{TABULA} and the average indoor temperature throughout the year is assumed to be constant at 20°C in the heat demand calculations, the Modelica model sets the indoor set temperature to 20°C throughout the year and selects the two time periods from the beginning of the year to May 10, and from October 1 to the end of the year (a total of 222 days) as the heating season, based on the temperature profile used (see \Cref{fig: Tout and Qsum}, top) to ensure the reliability of the validation.

\subsubsection{Set points of ventilator}
\label{subsubsec:Set points of Ventilator}
According to the regulatory standards or guidelines from several counties, carbon dioxide concentrations in buildings such as offices, schools, and residences should ideally be kept below 1000 to 1500 ppm \citep{hattori2022, health2021, yoshino2019, ashrae2022} and the indoor relative humidity should be kept between 30\% and 50\% \citep{usenv2024}. The airflow rates set in \ac{TABULA} and in the Modelica model are described in \Cref{subsec:Building envelope model} and in \Cref{tab: Buildings parameters}.

\subsubsection{Set points of air conditioner}
In order to maintain the indoor temperature at a set point throughout the year, it is not enough to simply control the indoor temperature through ventilation fans and radiators. For example, in summer, due to strong solar radiation, some buildings have indoor temperatures that are higher than the outside temperature without a cooling system. Therefore, for the need of validations and simulation calculations, air conditioners with heating and cooling control functions are added to the model.

It is worth mentioning that this air conditioning model can also provide or extract latent heat from the indoor air by detecting the humidity in the room in order to keep the indoor humidity at a set value. This function is not specifically discussed in the present paper since the main object of study is the sensible heat characteristics of the model.

\subsection{Weather data input}
Weather data is recorded at the Living Lab at \ac{KIT} Campus North and stored in a time series database with a Grafana (grafana.com) interface for further data analysis and export. The data comprises temperature, solar irradiation, humidity, wind speed, and direction, etc. \citep{hagenmeyer2016}. Outdoor temperature has an extremely important influence in the study of the heat demand of a building since the temperature difference between indoors and outdoors determines the amount of transmission heat transfer. In \ac{TABULA}, the average outdoor temperature throughout the year is assumed to be 10.512°C. Therefore, the temperature profile used in the present paper is also adjusted to this average value.

\subsection{Flexible parametrization of building models}
\label{subsec:Flexible Parametrization of House Models}
Since the model presented in the present paper is directed to a generic residential building model, the authors simulate all of the equipment that might be found in a real residential building, as described in \Cref{subsec:House equipment}. During this process, the authors include a control parameter (with a value of either 0 or 1) for each module to indicate the presence or absence of this component in the current building model.

\section{District heating system - model overview}
\label{sec:District Heating System}

The authors introduce a \ac{DH} simulation model, which is designed in Modelica using the AixLib model library \citep{müller2016} for main system components such as the heating plant, the pipes, and the substations. In order to facilitate co-simulation, the \ac{DH} model is built based on the open-loop modeling approach, which is described in more detail in \citep{fuchs2016}.

For the pipes in the \ac{DH} network, a static pipe model \textit{StaticPipe}, available from the AixLib library, is used to further reduce complexity and model size. The heating plant is modelled in a simplified way using the \textit{SourceIdeal} submodel, which represents an ideal heat source that prescribes a supply temperature to the mass flow and controls the supply pressure at the supply line depending on a pressure measurement at the most distanced building. For the demand model, which represents the primary side of the building substation, the \textit{VarTSupplyDp} submodel from the AixLib is used and adapted to the given characteristics in this study.

The \ac{DH} model is semi-automated generated using the energy network management tool \textit{uesgraphs} \citep{fuchs2016}, and the \ac{DH} system under investigation is represented as a graph structure in \textit{uesgraphs} using GIS data \citep{osm2024}. Once the graph structure in \textit{uesgraphs} has been enriched with information about the \ac{DH} system, the graph structure can be exported as a Modelica model using a template approach that allows specific parameterisation of the \ac{DH} components depending on the data implemented in the graph structure.

\section{Model and simulation coupling}
\label{sec:Model and Simulation Coupling}
This section introduces the interfaces for the coupling of the building and \ac{DH} grid model with a focus on the interface models on the building and the heat grid side.

\subsection{Interface on the building side}
\label{subsec:Interface on the Building Side}
This section discusses in detail the design ideas and principles of the \ac{HIU} model based on \Cref{fig: HIU} in the previous \Cref{subsec:Building and district heating grid interface}. The output signals of \ac{HIU}'s Modelica model are the buffer tank temperature $T_{buffer}$ and the total heat demand $\dot{Q}_{building}$ of the building. The input signals are the hot water supply or inflow temperature $T_{sup}$ from the \ac{DH} grid, the hot water flow rate $\Dot{m}$, and the requested hot water return temperature or outflow temperature $T_{ret,building}$ by the grid.

When the hot water from the heat grid flows into the \ac{HIU}, its heat is transferred to the water in the buffer tank in the heat exchanger (if the temperature of the incoming hot water $T_{sup}$ is higher than the water temperature of the buffer tank $T_{buffer}$), then it flows out of the \ac{HIU} at a temperature $T_{ret}$ higher than the water temperature in the buffer tank based on the physical characteristics of the heat exchanger and returns to the heating grid. During this process, $T_{buffer}$ should be maintained between 75°C and 85°C \citep{cibse2011}. In order to intuitively analyze the changes in the heat demand of a building throughout the year, a simple control strategy is adopted in the present paper, that is, the $T_{buffer}$ is controlled to be maintained at 80°C all year round. As mentioned in \Cref{subsec:House equipment}, one radiator model is connected to the buffer tank in \ac{HIU}. When the buffer tank reaches the appropriate temperature, hot water will be supplied from the buffer tank to the radiator.

\subsection{Interface on the district heating side}
\label{subsec:Interface on the District Heating Side}

\begin{figure}[htb]
    \centerline{\includegraphics[trim={0cm 0cm 0cm 1.5cm}, width=0.5\columnwidth]{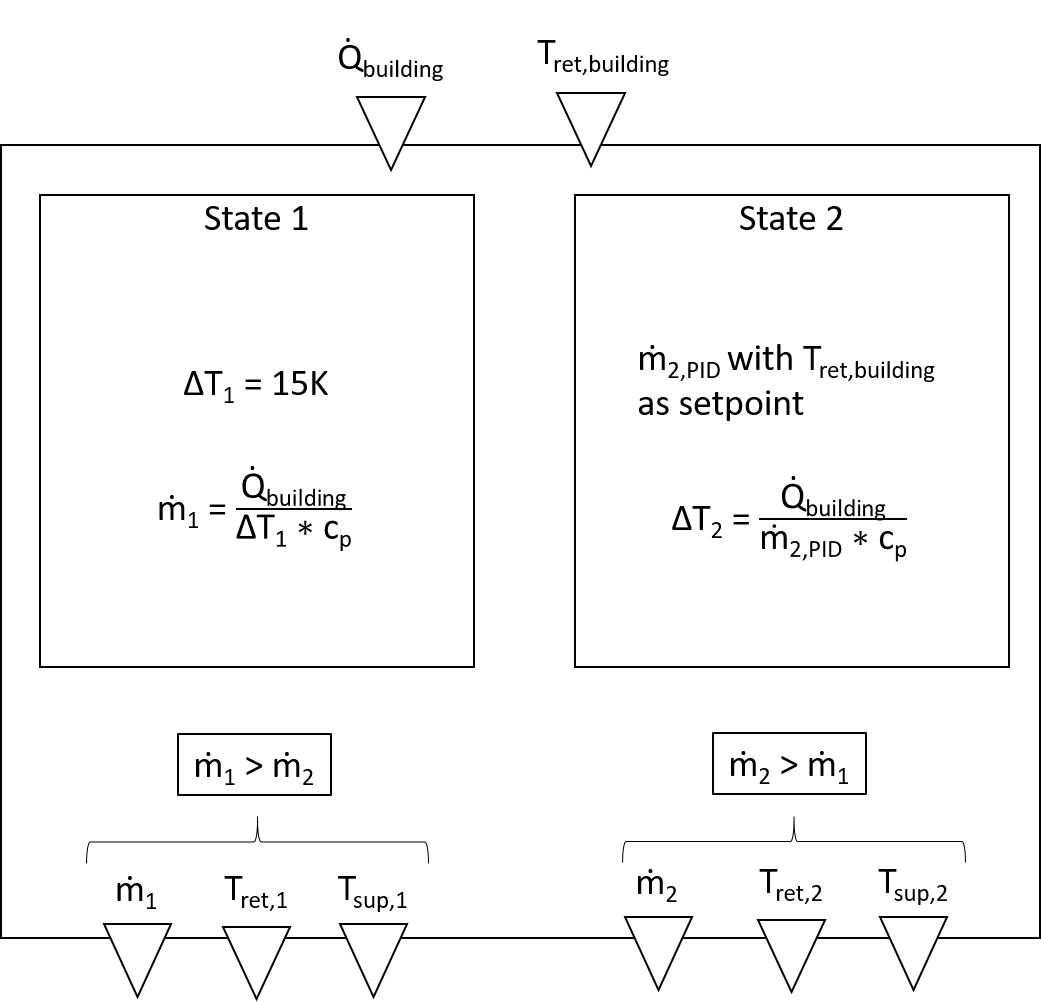}}
    \caption{Schematics of the \ac{DH} system substation model and its control system.}
    \label{fig: heating model substation control}
\end{figure}

The modified substation model based on the \textit{VarTSupplyDp} submodel of the Aixlib specifies the heating demand of the building as a required heat supply from the \ac{DH} system and it can take two states. State 1 is only controlled by the heat demand of the building and behaves like the Aixlib submodel \textit{VarTSupplyDp}. In this state, the mass flow $\Dot{m}_{1}$ with the given heat demand of the building $\dot{Q}_{building}$, a fixed temperature difference $\Delta T_{1}$ of $15K$, and the specific heat capacity $c_{p}$ are calculated. This calculation results in the required mass flow of the \ac{DH} system on the substation primary side to cover the heat demand of the building $\dot{Q}_{building}$. This state does not consider the buffer temperature level of the building. Therefore, state 2 is implemented and considers the required temperature level to supply the building sufficiently. If the supply temperature of the \ac{DH} system drops below the buffer temperature of the building heating system, control state 2 takes over. \Cref{fig: heating model substation control} shows both states of the substation and the control approach of state 2.

State 2 attempts to tackle this problem by controlling the mass flow $\Dot{m}_{2,PID}$ using a PID controller. The set point of the controller is represented by the buffer temperature of the building $T_{buffer}$. If the required temperature level is not reached by the \ac{DH} system temperatures due to high heat losses and, therefore, high-temperature drops in the \ac{DH} network, the mass flow must be increased to reach higher temperature levels at the substation. Thus, a higher mass flow $\Dot{m}_{2}$ results in a higher supply $T_{sup}$ and, depending on the temperature difference at the substation, in higher return temperatures. If the temperature level at the substation is sufficient, the required temperature difference $\Delta T_{2}$ is calculated based on the current mass flow $\Dot{m}_{2,PID}$ to cover the heat demand at the substation and to reach the requested return temperature $T_{ret,building}$ by the grid.

\subsection{Distributed co-simulation}
\label{subsec:Distributed Collaborative Co-Simulation}

Co-simulation \citep{gomes2018} through the \ac{FMI} interface standard \citep{fmi2024} allows multiple subsystems to operate together in the same simulation environment. Since the simulation algorithms of each subsystem are integrated into a single system and exchanged variables, the \ac{FMI} protocol greatly reduces the divide between different subject models, providing great convenience for large-scale multi-subject energy system co-simulation \citep{alfalouji2023}. The \ac{FMU} export mechanism of Dymola, which incorporates a multi-physics model together with the executable simulation code and the interface definition, enables the use of sophisticated co-simulation frameworks as the eASiMOV-eCoSim (energy system Co-Simulation) \citep{cakmak2019}. The co-simulation environment allows for geographically distributed co-simulation with support for \ac{FMU}s \citep{kocher2024}, generated by Dymola or Simulink or any other capable software, but also simulators provided as Python or Matlab source code. This feature is important since it allows more flexibility in the simulator coupling and enhances the possibilities for simulator design by code.

\begin{figure}[htb]
    \centerline{\includegraphics[width=0.45\columnwidth]{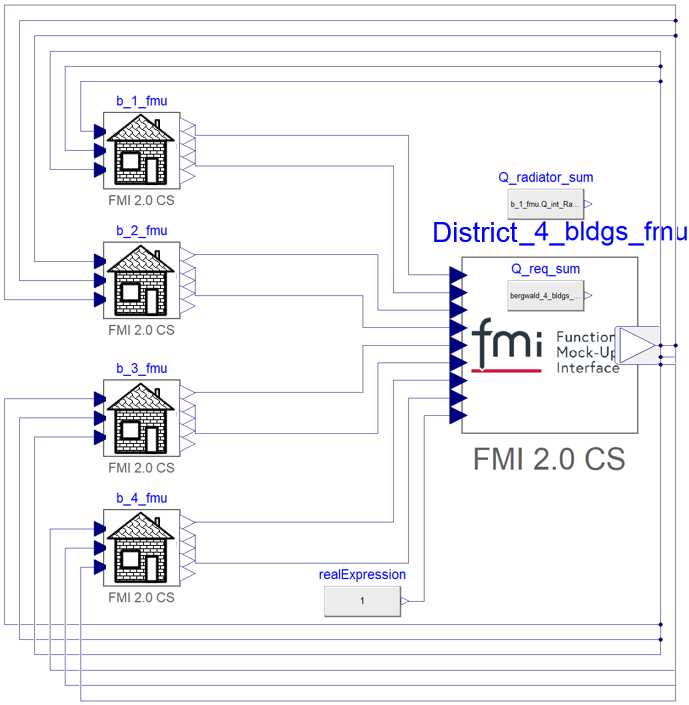}}
    \caption{Co-Simulation setup of the tested energy district grid comprising buildings with different properties and heat grid as \ac{FMU}s in \textit{Dymola}.}
    \label{fig: CoSim Dymola}
\end{figure}

For the present study, all multi-physics building models are exported as \ac{FMU}s, whose dependencies are shown in \Cref{fig: CoSim Dymola} for the present study.
The co-simulation is composed finally by importing and interconnecting the \ac{FMU} blocks and assigning the respective data exchange ports in the eCoSim graphical editor \citep{cakmak2019}. Additionally, all simulators (\ac{FMU}s) are assigned for parallel and sequential execution by insertion of logical delay blocks to achieve correct simulation synchronization. In this setup, a loose coupling of the simulators -- \ac{DH} model at \ac{FZJ} and building models at \ac{KIT} -- is used, whereby the master at \ac{KIT} orchestrates the co-simulation.

\section{Evaluation}
\label{sec:Evaluation}

In this chapter, the authors first present the heat demand simulation results of the developed novel building model under the configurations mentioned above and compare the simulation results with the reference values given in \ac{TABULA}. Further on, the authors will show the operation of the interface (i.e., the \ac{HIU}) between the building and the heat grid, and finally the simulation results will be analyzed and discussed.

The simulation model is shown in \Cref{fig: CoSim Dymola}, which contains four \ac{FMU} modules representing the building model (left side) and one \ac{FMU} module representing the heat grid (right side). The reason for selecting only four buildings as research objects is, that the present paper focuses on the analysis of model effectiveness. In fact, the model can perform co-simulation on all building types from TABULA with unlimited number of buildings. The start and end times of the simulation are set to the first day of one year and the first day of the second year, i.e., the total duration of the simulation is 1 year with a time resolution of 15 minutes. The runtime for the simulation is 1.57e+03 seconds, i.e., about 26 minutes and 10 seconds in Dymola with Intel\textsuperscript{\textregistered} Core\texttrademark\ i7-1255U CPU@1.70GHz and 16GB installed RAM. 
Simulation times are in average 4 times higher for geographically distributed co-simulation between \ac{FZJ} and \ac{KIT} due to simulator communication and data exchange via the Internet.

\subsection{Validation of building models}
\label{subsec:Validation of building models}

In \Cref{fig: 5R1C electrical scheme model} the building envelope model is shown. To make the Modelica model results comparable with the \ac{TABULA} reference values, it is necessary to select appropriate variables for observation and calculation. The names and meanings of the variables in \ac{TABULA} selected by the authors are shown in \Cref{tab: Observed variables}, as well as the variables' counterparts in Modelica models. In the table, $T_{air}$ is the average room temperature for the whole year; $Q_{ht,ven}$ represents the amount of heat lost from the building through air exchange, calculated in Modelica by the following equation:
\begin{equation}
Q_{ht,ven} = \frac{1}{3600000} \int \dot{m}_{air} c_{air} \Delta{T} \, dt
\end{equation}
where $\dot{m}_{air}$ is the air mass flow rate in $kg/s$ of ventilation, determined by the $n_{air,rate,red}$ of equation (12) described in \Cref{subsec:Building envelope model}, $c_{air} = 1012 J/(kg \cdot K)$, and $\Delta{T} = T_{out} - T_{air}$; $Q_{ht,tr}$ is the sum of the heat flow through two heat conductance elements $H_{win}$ and $H_{tra}$ (see \Cref{fig: 5R1C electrical scheme model}) throughout the year.

\begin{table*}[htb]
\centering
    \caption{Variables observed and the meaning of them}
    \setlength{\fboxrule}{0pt}
    \fbox{\includegraphics[width=\linewidth,trim={0.0cm 0.0cm 0.0cm 0.0cm},clip]{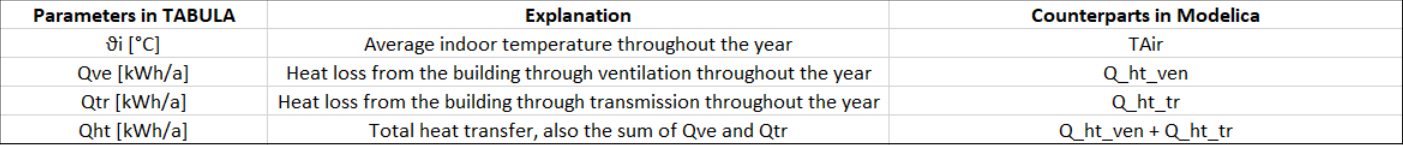}}
    \label{tab: Observed variables}
\end{table*}

The resulting calculations for the four buildings and the error data for each of the variables are shown in \Cref{tab: Validation results}.
As can be seen in the table, the four buildings have different heat transfer values due to different parameter configurations. The error values (in \%) between the simulation results and the \ac{TABULA} data are highlighted in red. In contrast, the error values for the total heat transfer are additionally marked with an orange background in the table. It can be summarized that the first and third building in the existing state have relatively high errors of about -6.5\% and -3.5\%, respectively, while the remaining two buildings in the usual refurbishment state have errors of less than 1\%. These results indicate that the error between the heat loss due to air exchange in the Modelica model and the corresponding data in \ac{TABULA} is minimal, not more than 1.3\%, so the main error comes from the transmission error. In addition, there is also a strong relationship between the total heat loss of a building and the temperature difference between indoors and outdoors. In terms of indoor temperature, it can be seen that Dymola's calculated output does not exactly match the set temperature (20°C). Since the \ac{HVAC} system is modeled using a fluid model (rather than an ideal model) in Dymola, the delayed behavior of the fluid causes the \ac{HVAC} system not to be completely accurate in controlling the temperature of the building. Even so, one can notice that the buildings with refurbishment have higher average temperatures, i.e., better insulation, and the total heat loss, especially heat transfer loss, is significantly reduced compared to the non-renovated buildings.

\begin{table*}[htb]
\centering
    \caption{Results: Comparison and error of Modelica calculating results with \ac{TABULA} data}
    \setlength{\fboxrule}{0pt}
    \fbox{\includegraphics[width=1\linewidth,trim={0.0cm 0.0cm 0.0cm 0.0cm},clip]{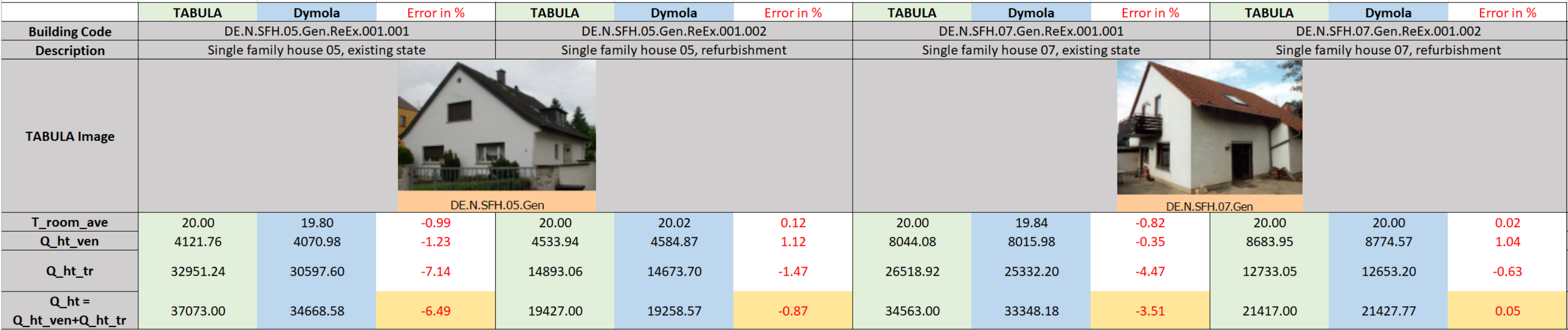}}
    \label{tab: Validation results}
\end{table*}

\subsection{Co-Simulation of buildings and heat grid}
\label{subsec:Co-Simulation of Buildings and Heat Grid}
In order to more accurately observe the effect of the parameter configuration of the building model on the heating demand of the building, the authors turned off the air conditioning model and performed another simulation with all inputs and parameters unchanged. It is worth noting that, as mentioned in \Cref{subsec:Control of building model}, air conditioning can be used not only to raise indoor temperatures as an auxiliary means of heating a building during cold outdoor temperatures but also to prevent indoor temperatures from becoming overheated in the summer due to the effects of sunlight, for example. If the air-conditioning component is removed and the heating grid is still available, then buildings in hotter summer climates should have higher average temperatures throughout the year in terms of the average temperature of the building.

As shown in \Cref{tab: NoAC}, without the cooling function of the air conditioners in the summer, the average indoor temperatures in all four buildings have increased. Especially for the renovated buildings with lower U-values, their temperatures (yellow background) are always higher than those of the non-renovated buildings, which is consistent with the practical experience.

\begin{table*}[htb]
\centering
    \caption{Changes in average indoor temperatures throughout the year after removal of AC}
    \setlength{\fboxrule}{0pt}
    \fbox{\includegraphics[width=\linewidth,trim={0.0cm 0.0cm 0.0cm 0.0cm},clip]{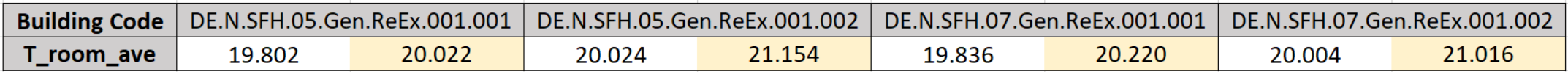}}
    \label{tab: NoAC}
\end{table*}

\begin{figure}[htb]
    \centerline{\includegraphics[width=0.7\columnwidth]{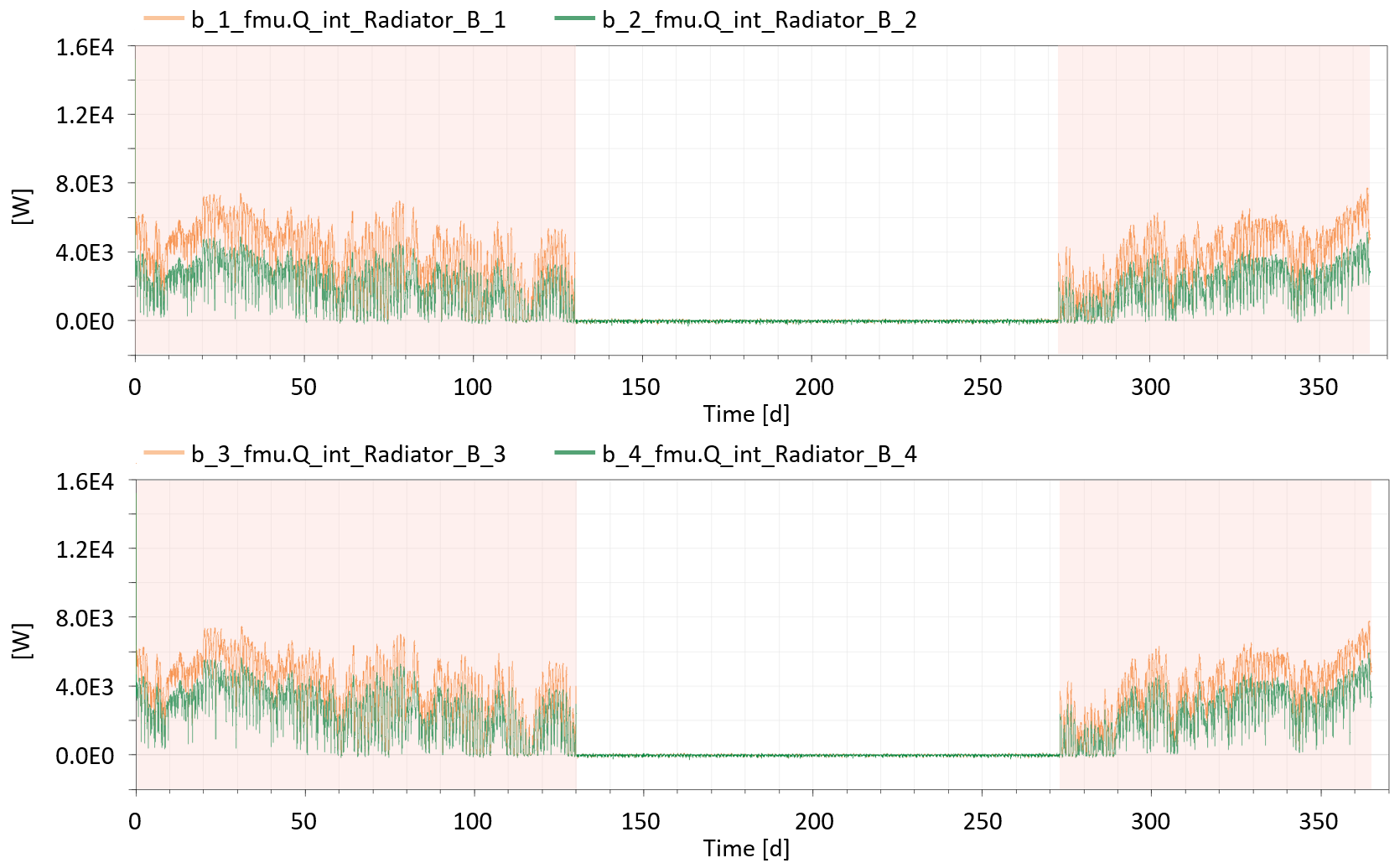}}
    \caption{Radiator heat power change curves for building SFH.05 (top) and building SFH.07 (bottom), orange indicates power in the existing state, while green indicates power after refurbishment.}
    \label{fig: Q radiator}
\end{figure}

\begin{figure}[htb]
    \centerline{\includegraphics[width=0.7\columnwidth]{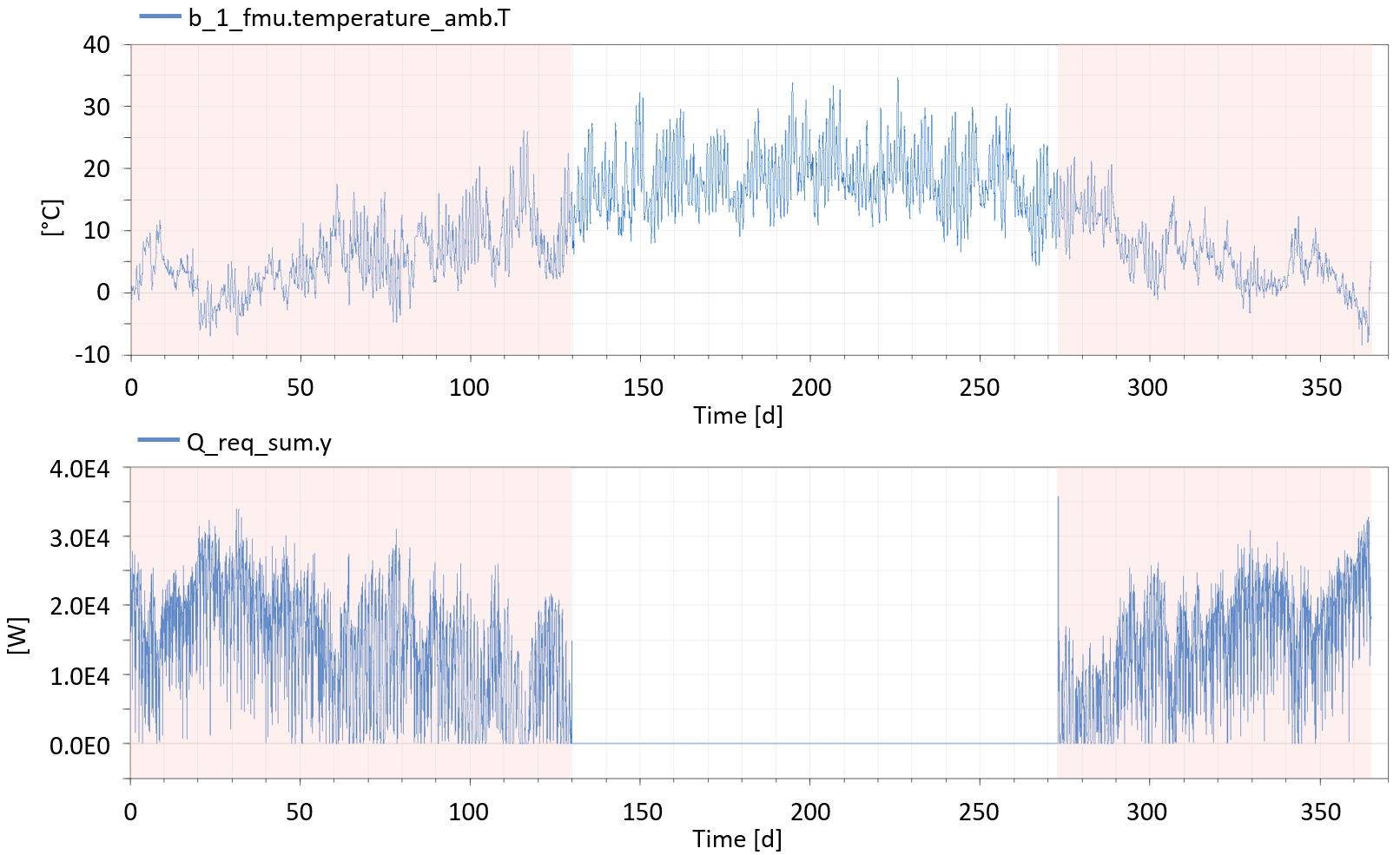}}
    \caption{Recorded outside temperature for the whole year (top) and the total provided heat power from the heat grid (bottom).}
    \label{fig: Tout and Qsum}
\end{figure}

In this simulation, apart from the incoming heat from sunlight and the operation of electrical appliances, the main heat of the building is provided by indoor radiators connected to the heat grid (or \ac{HIU}). The heat supplied to the room by the radiators in the four buildings is shown in \Cref{fig: Q radiator}. Also, the outdoor temperature throughout the year and the sum of the radiator power of the four buildings, but also the sum of heating power from the heat grid, are shown in \Cref{fig: Tout and Qsum}.

\begin{figure}[htb]
    \centerline{\includegraphics[width=0.7\columnwidth]{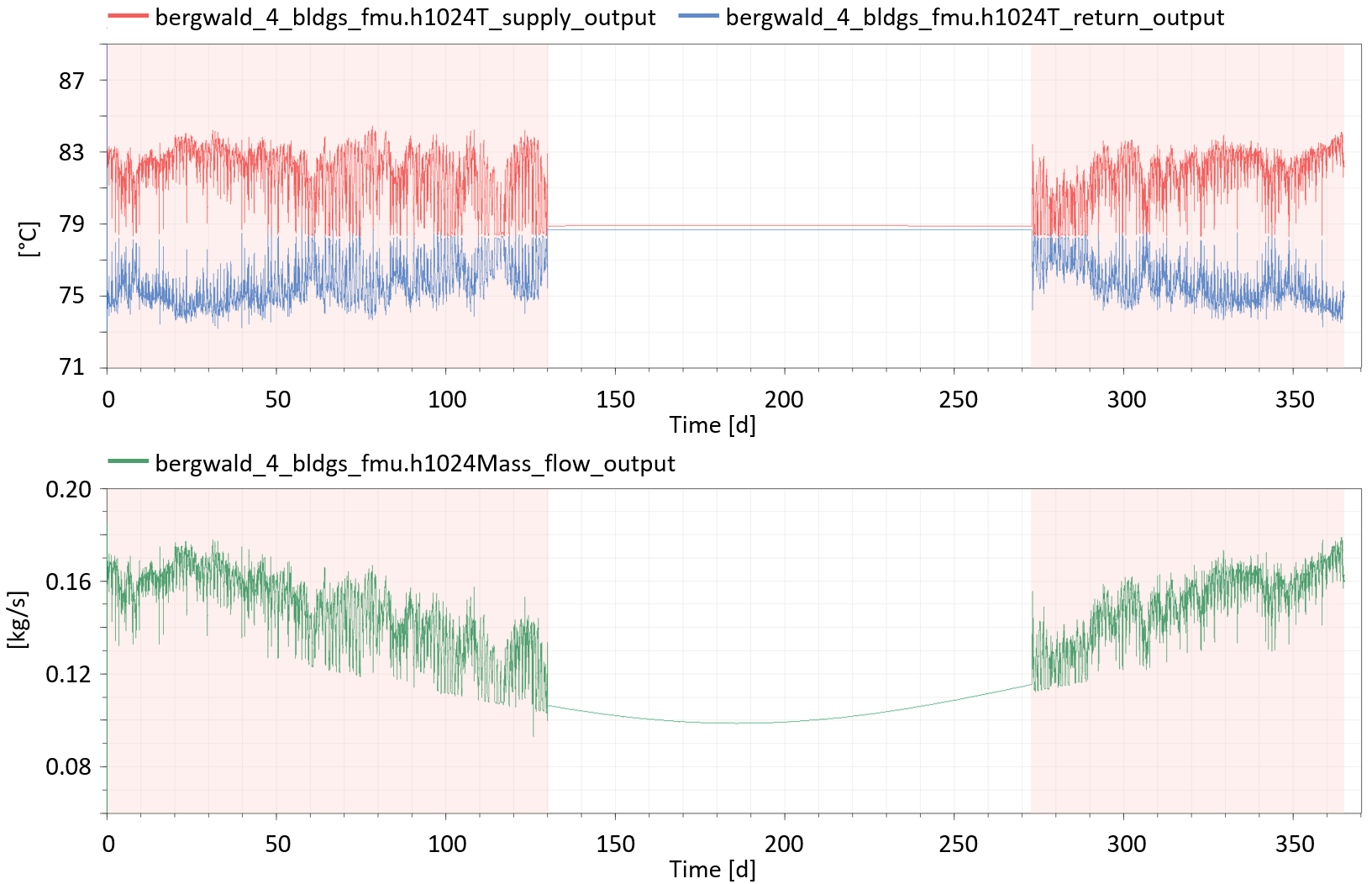}}
    \caption{Building DE.N.SFH.05.Gen.ReEx.001.001: Supply (red) and return (blue) temperature, and hot water mass flow rate (green) for the whole year.}
    \label{fig: Supply return temperature and mass flow rate of B1}
\end{figure}

It can be concluded from the results shown in \Cref{fig: Q radiator} and \Cref{fig: Tout and Qsum} that the insulation of a building has a significant impact on its heat demand, and the level of heat supply from the heat grid has a strong relationship with the outdoor temperature. When the outdoor temperature is low, the amount of heat lost from the building through air exchange and heat transmission increases, which results in the building needing higher thermal power to maintain a stable indoor temperature. The supply and return temperatures and the inlet mass flow rate of the \ac{DH} system of building DE.N.SFH.05.Gen.ReEx.001.001 are illustrated in \Cref{fig: Supply return temperature and mass flow rate of B1}. When the building's heating demand increases, the building sends this demand to the heat grid under the regulation of the \ac{HIU}. The required supply and return temperatures and the mass flow will be calculated and sent to the building model. By comparing the data in \Cref{fig: Tout and Qsum} and \Cref{fig: Supply return temperature and mass flow rate of B1} the authors observe that the temperature difference at the substation varies depending on the desired heat demand of the building.

Furthermore, it is important to note that the heat grid operates throughout the year, not just during the heating season. Nevertheless, as the authors suggest at \Cref{subsec:House equipment} to deactivate the \ac{DHW} system in the model, the heat grid will not provide any heat to the building during the non-heating season if the building does not have a demand for room heating. During this period, it can be observed from \Cref{fig: Supply return temperature and mass flow rate of B1} that the temperature of the supply and return line is essentially constant, with only minor differences. This is due to the constant desired return temperature from 80°C for temperature control of the buffer tank of the building as explained in \Cref{subsec:Interface on the District Heating Side}. In addition, this variation of the mass flow rate in the non-heating season is due to the changing ground temperature during this period. The results show that the control and modeling approaches explained in \Cref{subsec:Interface on the District Heating Side} work as expected and show the desired behavior. However, the parameters (i.e., the desired return temperature of the building) need to be adjusted for real-world applications.

\subsection{Discussion}
\label{subsec:Discussion}
In the previous section, the authors presented the results obtained during the validation of the model as well as the results of the co-simulation of the building and the heat grid. By means of the error values presented in \Cref{tab: Validation results}, it is possible to understand that the model presented in the paper obtains a high level of confidence. The main source of error is the transmission heat loss, i.e., the building envelope model needs to be improved.

Under the co-simulation of the four buildings and the heat grid, the results presented focus on the heating of the buildings as well as the operation of the heat grid during the heating season, since the domestic water consumption of the building inhabitants is not considered in the present paper. When the building heat demand is high, the building inlet and outlet temperature differences as well as the hot water flow rate are subsequently increased to ensure that the temperature in the building remains at the set value.

In the real world, however, there are many ways to control the fit between the heat grid and the building as well as the \ac{HIU}. In the present paper, only one relatively simple control method is shown, i.e., the temperature tracking control of the incoming water flow, and the buffer tank temperature is set to a constant value. In addition, the proposed model opens new possibilities for the study of topics such as the economic and environmental benefits of energy use, and facilitates the design of new \ac{HIU} controllers due to features such as flexibility and modularity of the model. With the utilization of the \ac{FMI} interface protocol, the controller will not be limited only to those designed by the Modelica language.

\section{Conclusion and outlook}
\label{sec:Conclusion and Outlook}

This paper presents a novel modular building model that is highly flexible, extendable, parameterizable, and configurable developed using the Modelica language. The model is validated for its effectiveness and accuracy by comparison with publicly available reference-building data in Typology Approach for Building Stock Energy Assessment (TABULA) project. Subsequently, the model is extended to four different parametric building models using the Functional Mock-up Interface (FMI) standard and integrated with a district heating (DH) grid model in a co-simulation environment/setup. This demonstration highlights the great potential of the proposed building model for co-simulating energy systems and affirms its ability to adapt and perform in complex simulation scenarios in view of sustainability.

With the increasing demand for digitalization and sector coupling, this paper provides new impulses for research into future energy systems. To further improve the applicability and performance of the present model, in future work, the integration of building models with additional energy grids and the development of capabilities will be highlighted, such as the reception, analysis, and prediction of real-time data to contribute to the development of advanced urban energy digital twins. This will enhance dynamic simulations to respond to changing environmental conditions and facilitate their implementation in real-world applications, ultimately driving innovation in energy-efficient building systems for sustainability.

\bibliographystyle{unsrtnat}
\bibliography{Validated_Multi-Physics_Building_Models_for_Multi-Energy_System_Analysis}

\end{document}